\documentclass[aps,pre,twocolumn,groupedaddress, showpacs]{revtex4-1}
\usepackage{lipsum}
\usepackage{epsfig}
\usepackage{amssymb}
\usepackage{amssymb,amsmath,amsthm,graphics}
\usepackage{color}
\usepackage[colorlinks=true,linkcolor=blue,citecolor=blue]{hyperref}

\begin{document}


\title{Bubble-like structures generated by activation of internal shape modes in two-dimensional sine-Gordon line solitons}


\author{M\'onica A. Garc\'ia-\~Nustes}
\email[]{monica.garcia@pucv.cl}
\affiliation{Instituto de F\'isica, Pontificia Universidad Cat\'olica de Valpara\'iso, Casilla 4059, Chile}
\author{Jorge A. Gonz\'alez} 
\email[]{jorgalbert3047@hotmail.com}
\affiliation{Department of physics, Florida International University, Miami, Florida 33199, United States}
\author{Juan F. Mar\'in}
\email[]{juanfmarinm@gmail.com}
\affiliation{Instituto de F\'isica, Pontificia Universidad Cat\'olica de Valpara\'iso, Casilla 4059, Chile}


\date{\today}

\begin{abstract}

Nonlinear waves that collide with localized defects exhibit complex behavior. Apart from reflection,
transmission, and annihilation of an incident wave, a local inhomogeneity can activate internal modes of solitons, producing many
impressive phenomena. In this work, we investigate a two-dimensional sine-Gordon model perturbed by a family of localized forces. We
observed the formation of bubble-like and drop-like structures due to local internal shape modes instabilities. We describe the formation
of such structures on the basis of a one-dimensional theory of activation of internal modes of sG solitons. An interpretation of the
observed phenomena, in the context of phase transitions theory, is given. Implications in Josephson junctions with a current dipole device
are discussed.

\end{abstract}


\maketitle



\section{Introduction}

Nonlinear wave propagation in inhomogeneous media is of great importance in many branches of the \hbox{natural} sciences. Most physical
systems are nonlinear and inhomogeneous. Therefore, realistic models usually require to introduce weak nonlinearities and inhomogeneities
on its description \cite{Nicolis1995}.

The propagation of solitons has been widely investigated in the literature due to its important applications \cite{Peyrard2004}. The
study of the disturbances on traveling solitons by the presence of spatial inhomogeneities, like
localized defects or boundary interphase walls, are of great interest in physics and biophysics \cite{Ivancevic2013, Aslanidi1999, Aslanidi1999,
Mornev2000, Mornev2000}. Depending on the geometrical size and shape of such inhomogeneities, there can be annihilation or reflection of an
incident nonlinear wave. These disturbances on the propagation regime may have important consequences for the behavior of the system \cite{Mornev2000,
Ivancevic2013}. For instance, propagation of fluxons in Josephson junctions (JJ's) may be affected by spatial heterogeneities in the junction
created by a current dipole device \cite{Barone1982, Ustinov2002, Malomed2004}. However, transmission, reflection, and annihilation are far from being the only possible
phenomena that may occur when solitary waves collide with localized spatial inhomogeneities.

The particle-like behavior of solitons is a well-known property \cite{Peyrard2004, Remoissenet2013}. Moreover, it has been proved that solitons
behave more as extended objects rather than point-like particles.  The activation of internal modes under the action of a family of
topologically equivalent inhomogeneous forces was proved analytically for $\phi^4$ and sine-Gordon (sG) models \cite{Gonzalez1992,
Gonzalez2002}. Such internal modes can lose stability, leading to very exciting phenomena as breaking of solitons, the creation of
kink-antikink pairs and formation of multikinks as the two-kink soliton\cite{Gonzalez2003, Gonzalez2006, GarciaNustes2012}. 

The sG model has been applied in many branches of physics as particle physics and condensed matter theory. The model can describe many
phenomena in solid state physics as domain walls in ferromagnets and fluxons in long JJ's \cite{Gonzalez2008,
Malomed2004, Barone1982}. In high energy physics and cosmology, topological defects in Klein-Gordon (KG) systems are relevant in brane world
scenarios and to describe phase transitions in the early universe \cite{Kibble1980, Gani1999, Gani2014, Gani2015, deBrito2014a, deBrito2014b,
Galiev2015}.  

In this work, we investigate the effect on the internal structure, and therefore on the dynamics,
of a two-dimensional sG line solitons due to space-localized inhomogeneities. The study is based on an analytical model of
solitons externally excited by a family of topologically equivalent forces in one dimension \cite{Gonzalez2002, Gonzalez2003}. We show
that the soliton dynamics is highly enriched by the interplay between the activation of internal modes and the kink-antikink interactions.
Shape modes instabilities lead to the formation of stable bubbles and drops. These are sustained by the presence of
the inhomogeneity itself, being unstable otherwise. We also show that two-kink solutions\cite{GarciaNustes2012, Bazeia2003}
can be also formed in two-dimensional systems, creating stable {\it two-kink bubbles}.
The remainder of this paper is organized as follows.
In section \ref{Sec:2} we give a brief overview of the analytical theory of the activation of internal shape modes in one-dimensional sG
solitons. We also introduce the two-dimensional model of the system and the numerical methods employed. In section \ref{Sec:3}
we show and discuss the results from numerical simulations. The robustness of the observed phenomena is discussed in
section \ref{Sec:4}, and an interpretation of our results in the context of phase transitions is given. Finally, in section \ref{Sec:6} we present
our concluding remarks.

\section{Two-dimensional model and the one-dimensional theory of activation of shape modes\label{Sec:2}}

Consider the two-dimensional perturbed sG equation
\begin{equation}
 \label{Eq01}
 \partial_{tt}\phi(\mathbf{r},t)=\nabla^2\phi(\mathbf{r},t)-\sin\phi(\mathbf{r},t)-\gamma\partial_t\phi(\mathbf{r}, t)+F(\mathbf{r}),
\end{equation}
where $\mathbf{r}=(x,y)$, $\gamma$ is a linear-damping coefficient and $F$ is an external space-dependent force. Line solitons, pulsons,
ring solitons, and breathers are some examples of the many soliton solutions of equation \eqref{Eq01} for
$F(\mathbf{r})=0$ \cite{Christiansen1981, Martinov1992, Tamga1995}.

In one dimensional systems, when the inhomogeneous force is given by \hbox{$F(x) = \pm 2(B^2 -1)\mbox{sinh}(Bx)\mbox{sech}^2(Bx)$}, the
perturbed sG equation has the exact solutions $\phi_{\pm}(x)=4\arctan\exp\left(\pm Bx\right)$ \cite{Gonzalez2002}.
Here, $\phi_+$ and $\phi_-$ reads for the kink and antikink, respectively. $F(x)$ is an antisymmetric spatial
function  that vanishes exponentially for $x\to\pm\infty$ and has a single zero at $x_*=0$. It is known that the zeros of the external
force act as equilibrium positions for the soliton motion \cite{Gonzalez1992}. Thus, $x_*$ is an equilibrium position. In particular, if
$F(x)$ possesses only one zero ($F(x_{*}) = 0$) and $\partial F(x)/\partial x\big|_{x=x_{*}} >0$,  then  the point $x = x_{*}$ is a stable
(unstable) equilibrium position for the kink (antikink). Otherwise, if $\partial F(x)/\partial x\big|_{x=x_{*}} <0$, the equilibrium
position $x = x_{*}$ is unstable (stable) for the kink (antikink). Thus, $B$ is a parameter that controls the extension of the force, its
maximum amplitude, and the sign of its derivative at $x_*$.

For the following discussion we consider the case $\phi(x,0)=4\arctan\exp\left(Bx\right)$ and
$F(x) = 2(B^2 -1)\mbox{sinh}(Bx)\mbox{sech}^2(Bx)$.
For this force function, the stability analysis for kink internal modes can be
solved exactly \cite{Gonzalez2002, Gonzalez2003}. For $B^2>1$, $F(x)$ has a positive slope around $x_*$, the translational mode is stable
and there are no internal shape modes. In this regime, the soliton is trapped at the stable equilibrium position $x_{*}$. If $1/3<B^2<1$,
$F(x)$ has a negative slope around $x_*$. The translational mode of the kink is unstable. Still, no internal modes are present. The
soliton moves from its equilibrium position without shape deformations. For $1/6<B^2<1/3$, the first stable internal shape mode arises,
while for $B^2<1/6$ many other internal modes can appear. For $B^2<2/[\Lambda_*(\Lambda_*+1)]$, where $\Lambda_*=(5+\sqrt{17})/2$, the
first internal mode becomes unstable, leading to a soliton breakup and shape instabilities \cite{Gonzalez1992,Gonzalez2002, Gonzalez2003}.

\begin{figure}
\begin{center}
 \scalebox{0.36}{\includegraphics{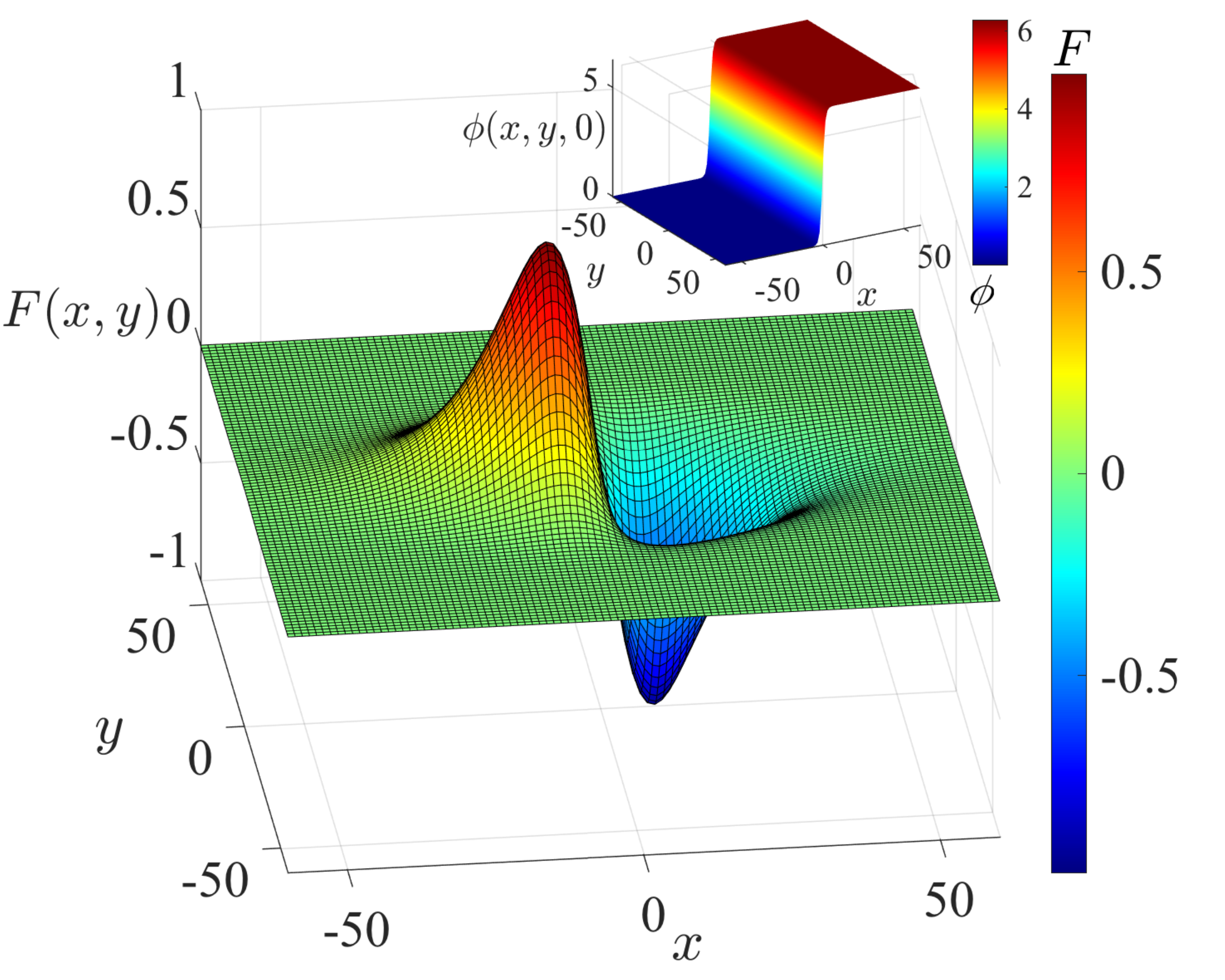}}
 \caption{(Color online) Inhomogeneous force of equation \ref{Eq02} for $B=0.1$ and $\sigma=15$. The inset show the static line
 soliton used as initial condition. \label{fig01}}
 \end{center}
\end{figure}

Following up the one-dimensional case, we will consider only line soliton solutions for the two-dimensional system, Eq.\eqref{Eq01}. To
control the spreading of the inhomogeneous force in the y-direction we introduce a smooth decaying function. Consequently, the internal
modes of the soliton will be activated locally, i.e., in a certain region of the kink front. Any rapidly decaying function can be used for
this purpose, e.g. a Gaussian function, a superposition of Heaviside functions, or some spline polynomial. In our model, we have used a
Gaussian function, since it is a physically realizable force. Therefore, the inhomogeneous force $F(x,y)$ is given by
\begin{equation}
\label{Eq02}
F(x, y)=\pm2(B^2 -1)\frac{\mbox{sinh}(Bx)}{\mbox{cosh}^2(Bx)}e^{-y^2/\sigma^2},
\end{equation}
where $B$ and $\sigma$ acts as control parameters. Given that $\sigma$  is a measure of the Gaussian width, it plays a controlling role
of the local stability of the front. Force \eqref{Eq02} is displayed in Figure \ref{fig01}. The perturbation of fluxons by dipole
current devices in JJ's is a typical example of a strong and local external influence with the same topological form as expression
\eqref{Eq02} \cite{Malomed2004}. This forcing term may be also relevant for 1D models of biological molecules, like DNA chains, where 1D sG systems have
been considered as a suitable model to explain the formation of open states or {\it bubbles} in the double helix \cite{Polozov1988, Yakushevich2004,
Grinevich2015}. Thus, results obtained in this work can be generalized to other topologically equivalent systems \cite{Gonzalez2002}.

 \begin{figure*}[t]
\centering
\includegraphics[width=1.0\columnwidth]{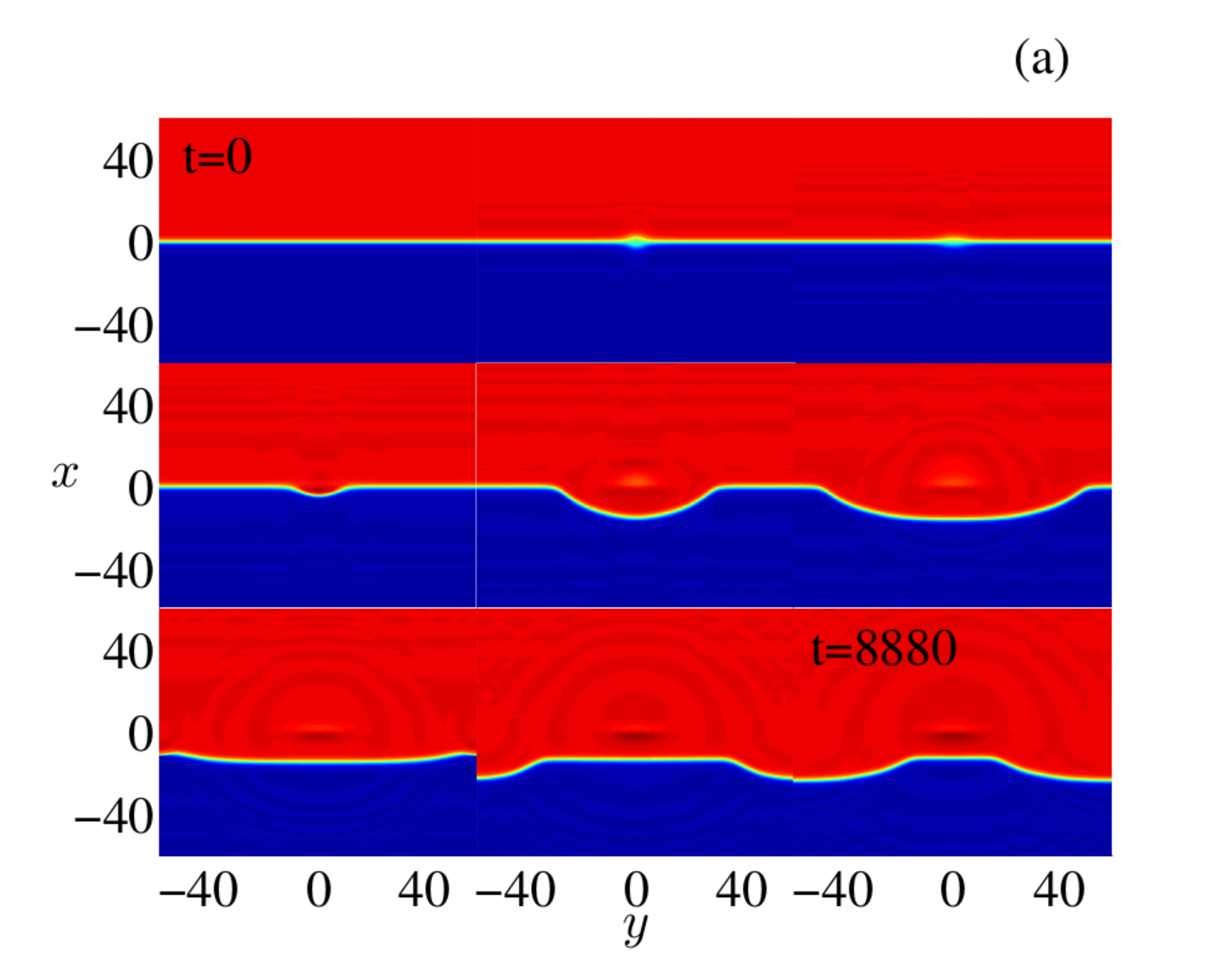}\quad\includegraphics[width=0.98\columnwidth]{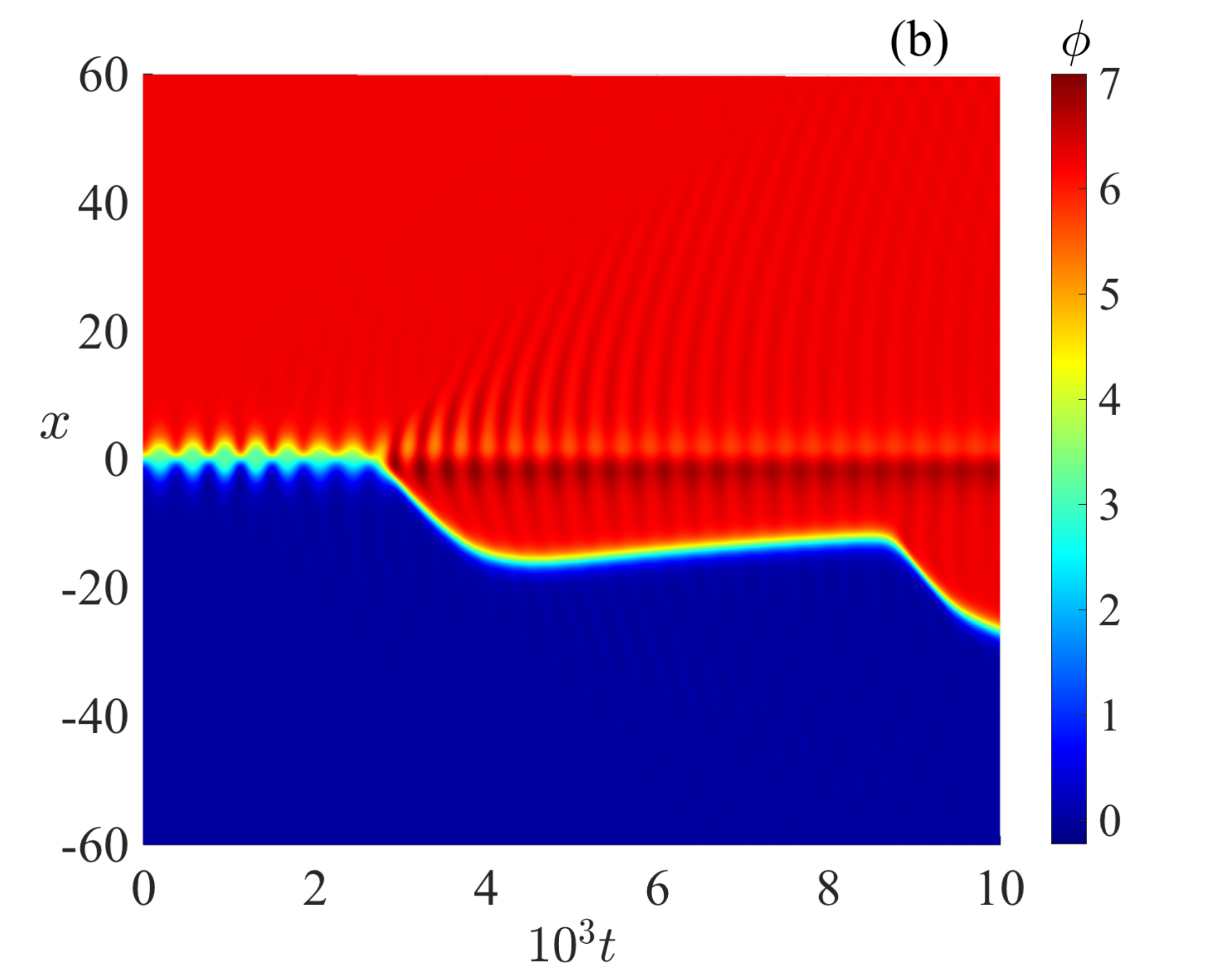}
\caption{\label{fig02} (Color online) Locally unstable soliton  ($B^2=0.5$ and $\sigma=15$). (a) Snapshots from numerical simulations, taken each
 1110 time iterations. (b) Time evolution of the $x$-profile of the soliton ($y=0$). The value of the field $\phi$ is indicated in
 color scale.}
\end{figure*}

In the present paper, all numerical simulations are performed using a kink as the initial condition, and the
perturbation $F(x, y)=2(B^2 -1)\mbox{sinh}(Bx)\mbox{sech}^2(Bx)e^{-y^2/\sigma^2}$. However, recall that the coefficient $2(B^2-1)$
becomes negative for $B^2<1$. This change will lead to the instabilities that we have discussed above.
We have used finite differences of second order of accuracy to numerically solve equation \eqref{Eq01} with no-flux boundary conditions.
For the explicit difference schemes, we have used a regular mesh of size $l\times l$, with $l=1200$. The space intervals are
$\Delta x=\Delta y=0.1$, and the time increment is \hbox{$\Delta t=0.02$}. The initial conditions are given by
$\phi(x,y,0)=4\arctan\,\exp(x/L)$ and $\partial_t{\phi}(x,y,0)=0$, where $L$ is an arbitrary parameter. This represents a stationary kink
located at the equilibrium point $x_*$ (see inset  Fig.\ref{fig01}). Under these conditions, we avoid any scattering effect, focusing our
attention on the dynamics of the soliton internal structure. Notwithstanding, any stationary soliton located near the stable equilibrium
point $x_*$ can be attracted and trapped by the inhomogeneity, setting up an initial condition as well. The damping term in equation \eqref{Eq01} is
introduced to dissipate the phononic excitations emitted by the mesh.

\section{Numerical simulations\label{Sec:3}}

In this section, we summarize the main results obtained from numerical simulations. As previously stated, $B$ and $\sigma$ act as control
parameters. Thus, our parameter space is $\{B,\sigma\}$. It is expected that for a value of $\sigma$ large enough, we recover the
one-dimensional results by varying $B$. For smaller values of $\sigma$, however, activation of internal modes will be only local.

\subsection{Stability of the translational mode}
 
To check the numerical method, we started our simulations setting the following parameter values, $B=1.2$ and $\sigma_{max}=15$. This
value reproduces an extended enough force in comparison with the mesh size $l$ and it represents our  $\sigma$ upper limit. According to
the one-dimensional theory, for $B^2>1$, it is expected that the translational mode to be stable, whereas no shape internal modes are
present. Indeed, the line soliton translational mode is stable for $B=1.2$. No front motion is observed.

Next, we decrease the value of $B$. According to the 1D theory, for $1/3 < B^2 < 1$ the translational mode is locally unstable.

Two-dimensional numerical simulations show that for $1/3 < B^2 < 1$, the localized force \emph{always} destabilizes the translational mode
of the soliton, despite the value of $\sigma$. This is a surprising result since one would expect that a sufficiently localized
perturbation could not destabilize an extended object. The lattice coupling would prevent the kink front motion if the perturbation
is sufficiently local. However, this is not the case.

\begin{figure*}[t]
\centering
\includegraphics[width=1.0\columnwidth]{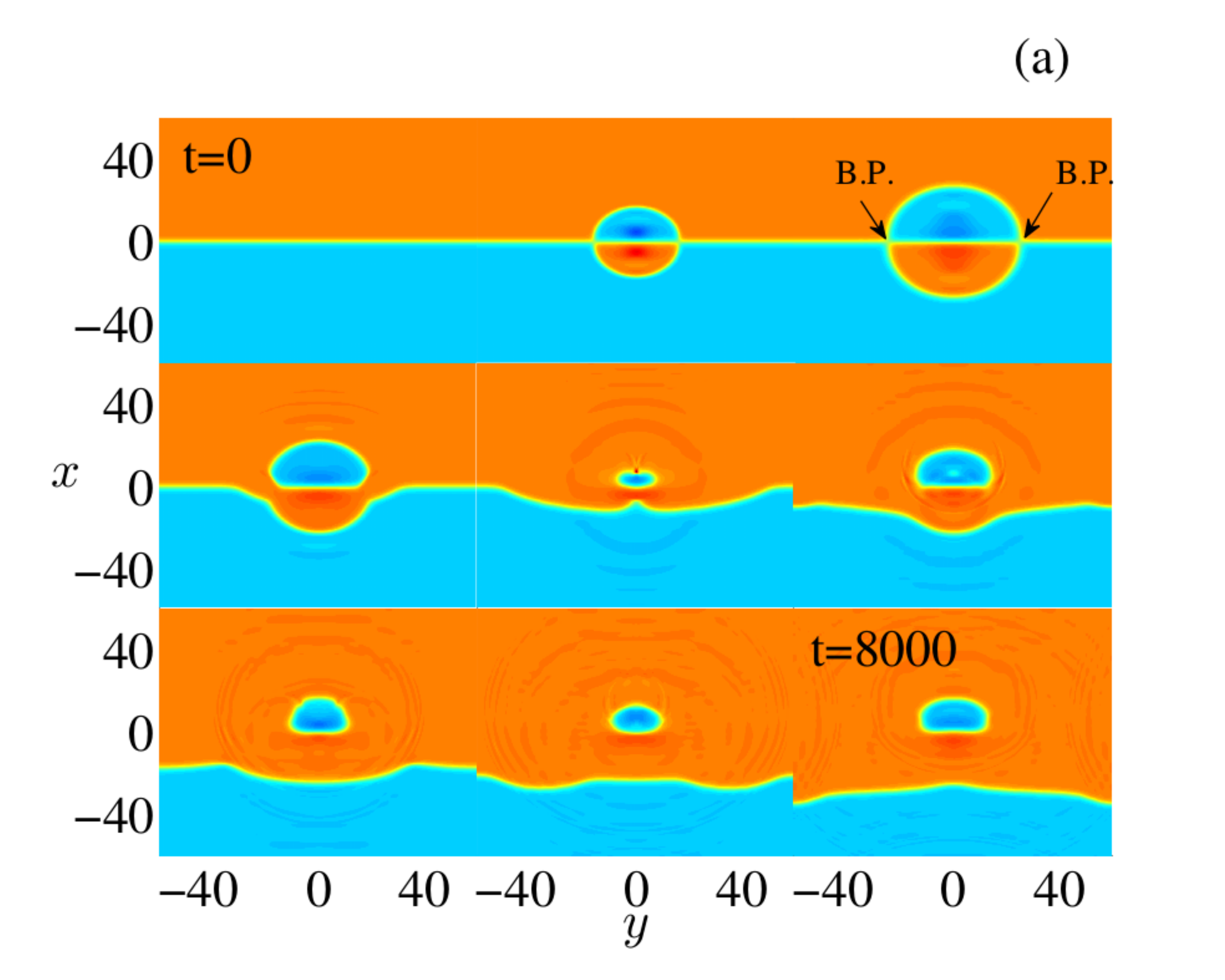}\;\includegraphics[width=0.97\columnwidth]{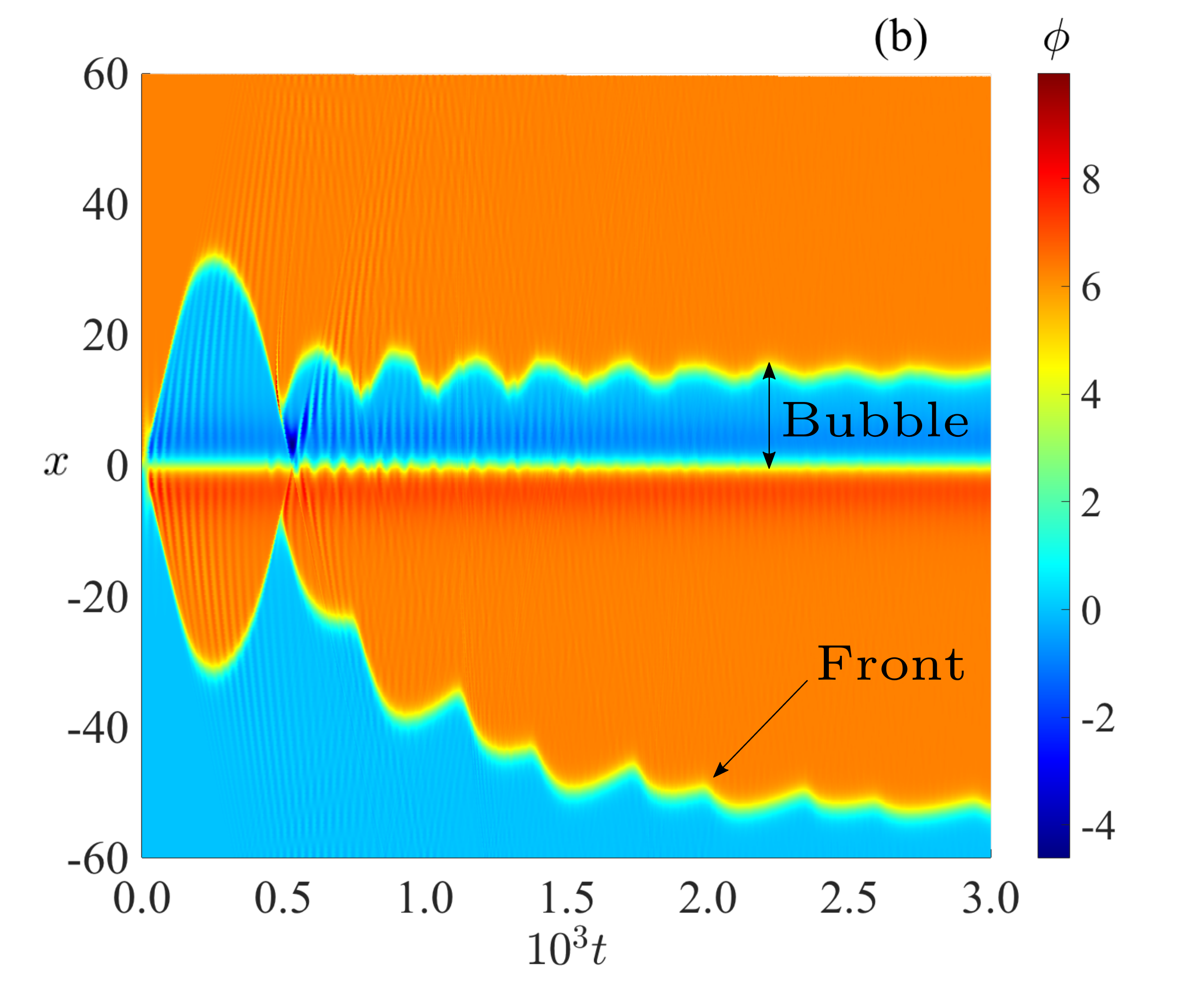}\\
\includegraphics[width=0.95\columnwidth]{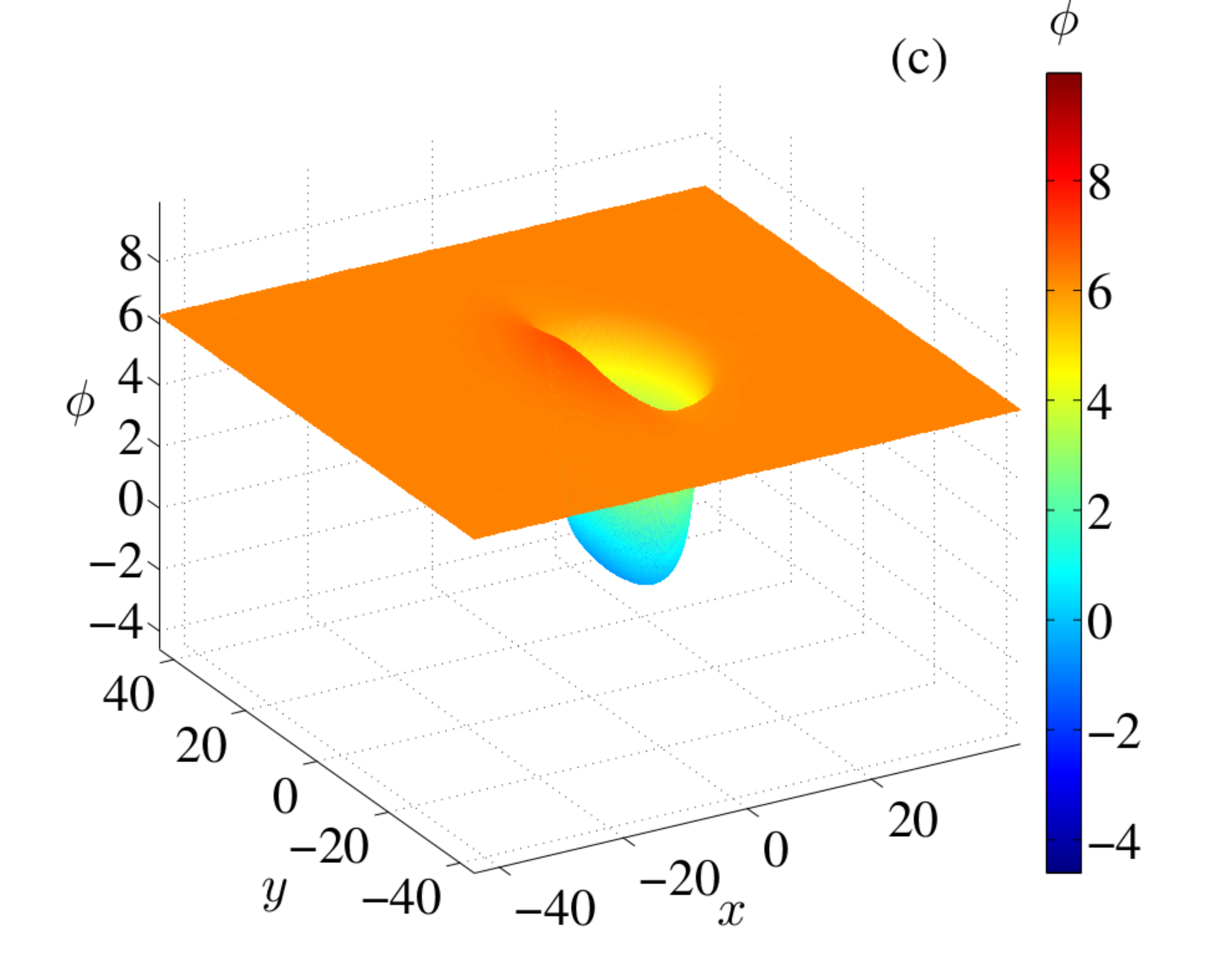}
\caption{\label{fig03} (Color online) Bubble formation for $B=0.27$ and $\sigma=15$. (a)  Snapshots from numerical simulations, taken each 1000 time
 iterations. (b)  Time evolution of the $x$-profile of the soliton for $y=0$. (c) Bubble-like structure after 30000 time iterations.
 The value of the field $\phi$ is indicated in color scale.}
\end{figure*}

In figure \ref{fig02} we show the results obtained for $B^2=0.5$ and $\sigma=15$. The soliton translational mode is locally
unstable around the origin, as expected from the one-dimensional theory. Since the force is localized in space, the kink front turns
unstable locally. In figure \ref{fig02}(a) we see that this local instability travels in the $y$-direction with a certain velocity,
destabilizing the whole kink front progressively. The per\-tur\-ba\-tion eventually reflects on the boundaries, and trav\-els backward.
In figure \ref{fig02}(b), we display the $x$-profile for $y=0$. We can observe that the line soliton core experiences a small
oscillation before the whole front begins to move. Such oscillation, which amplitude is decaying in time, is produced by an interaction
between the local destabilization and the lattice. Eventually, the local instability overcomes the lattice coupling coefficient and the
front starts to move after around $2000$ time iterations. Figure  \ref{fig02}(b) also reveals a curvature effect, exhibited by the small
return of the front to the origin. Such effect will be described more extensively in the subsequent section. The last bursting is produced
by the reflection of the perturbation in the walls.

\subsection{Bubble and drop formation}

\begin{figure}
\begin{center}
 \scalebox{0.7}{\includegraphics{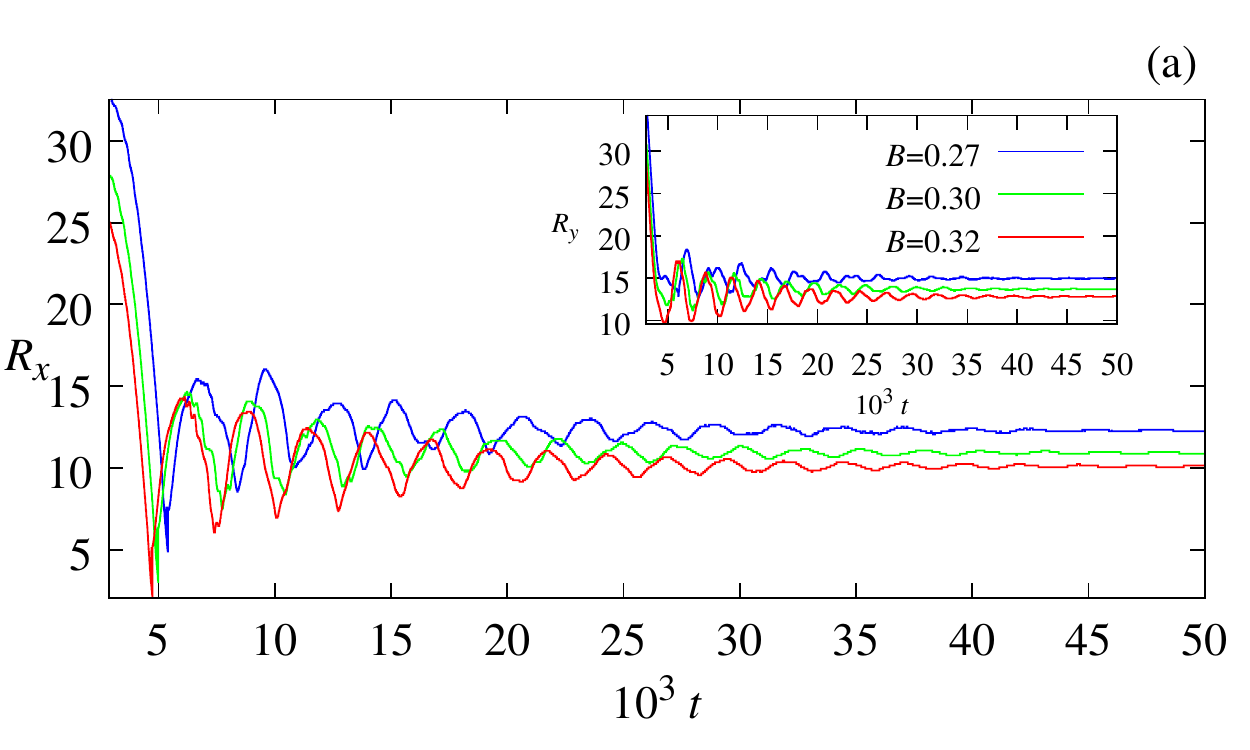}}\\
 \scalebox{0.7}{\includegraphics{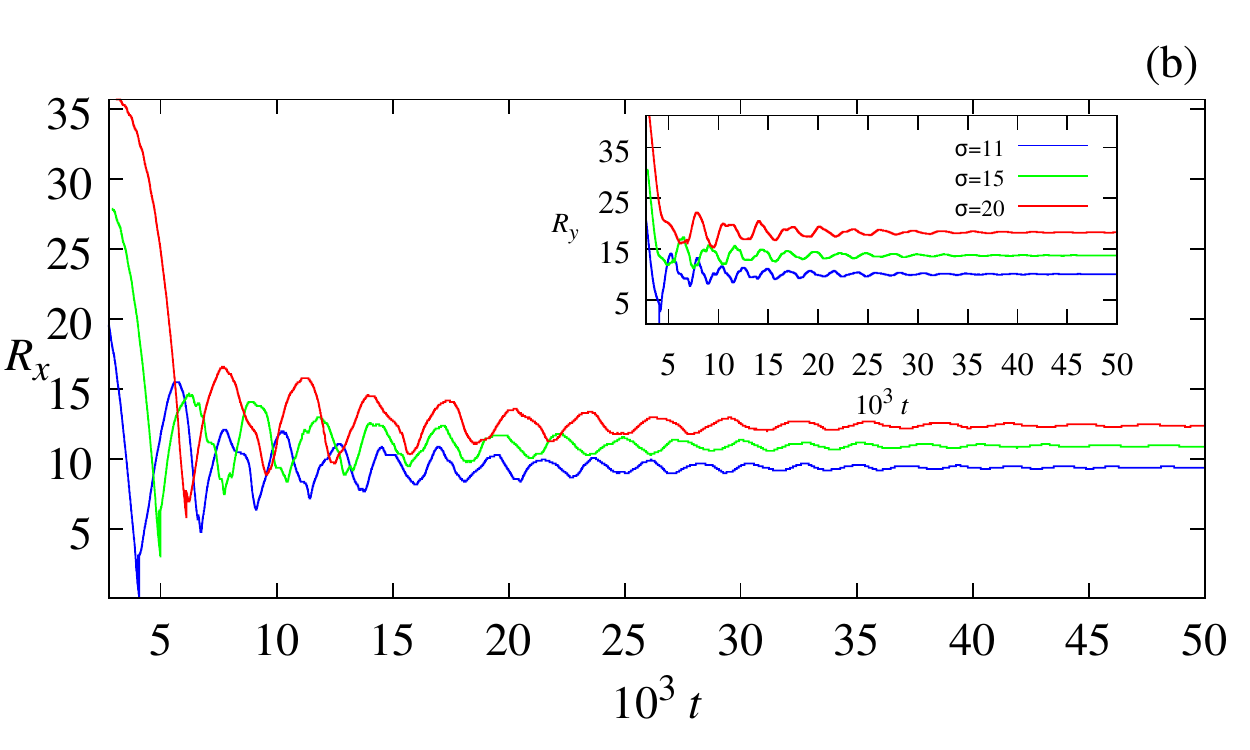}}
 \caption{(Color online) Evolution of the semi-axis of bubbles as a function of time for different values of parameters $B$ and $\sigma$. (a)
 Evolution of $R_x$ and $R_y$ (inset) for $\sigma=15$.
 (b) Evolution of $R_x$ and $R_y$ (inset) for $B=0.27$. \label{fig04}}
 \end{center}
\end{figure}

For $1/6<B^2<1/3$, the stable first internal shape mode arises, which becomes unstable for $B^2<2/[\Lambda_*(\Lambda_*+1)]$, where
$\Lambda_*=(5+\sqrt{17})/2$. Consequently, for $0<\sigma\leq \sigma_{max}$, the line soliton will exhibit more complex phenomena due to
the local destabilization of an internal shape mode. Effectively, the local instability of the shape mode leads to a local break-up of the
front. This initial breaking travels through the lattice, developing finally a stable \emph{bubble-like structure}. In figure \ref{fig03}(a), we
can observe the formation of a stable bubble-like structure for $B=0.27$ and $\sigma=15$. Initially,  we observe the formation of an
elliptic-like structure that grows around the inhomogeneity. The initial small oscillation is not present because the instability is
strong enough to overcome rapidly the lattice coupling coefficient. In this scenario, the curvature of the structure plays an important
role in the dynamics. The radial velocity decreases such that a maximum size of the elliptic structure is reached, and then the
structure collapses backward. This effect is known in the literature as the \emph{return-effect} \cite{Christiansen1978}, and is produced
entirely by the curvature of the elliptic front. During this collapse, the elliptic structure separates from the front in two
breaking points (BP) showed in figure \ref{fig03}(a). 

\begin{figure*}[t]
\centering
\scalebox{0.35}{\includegraphics{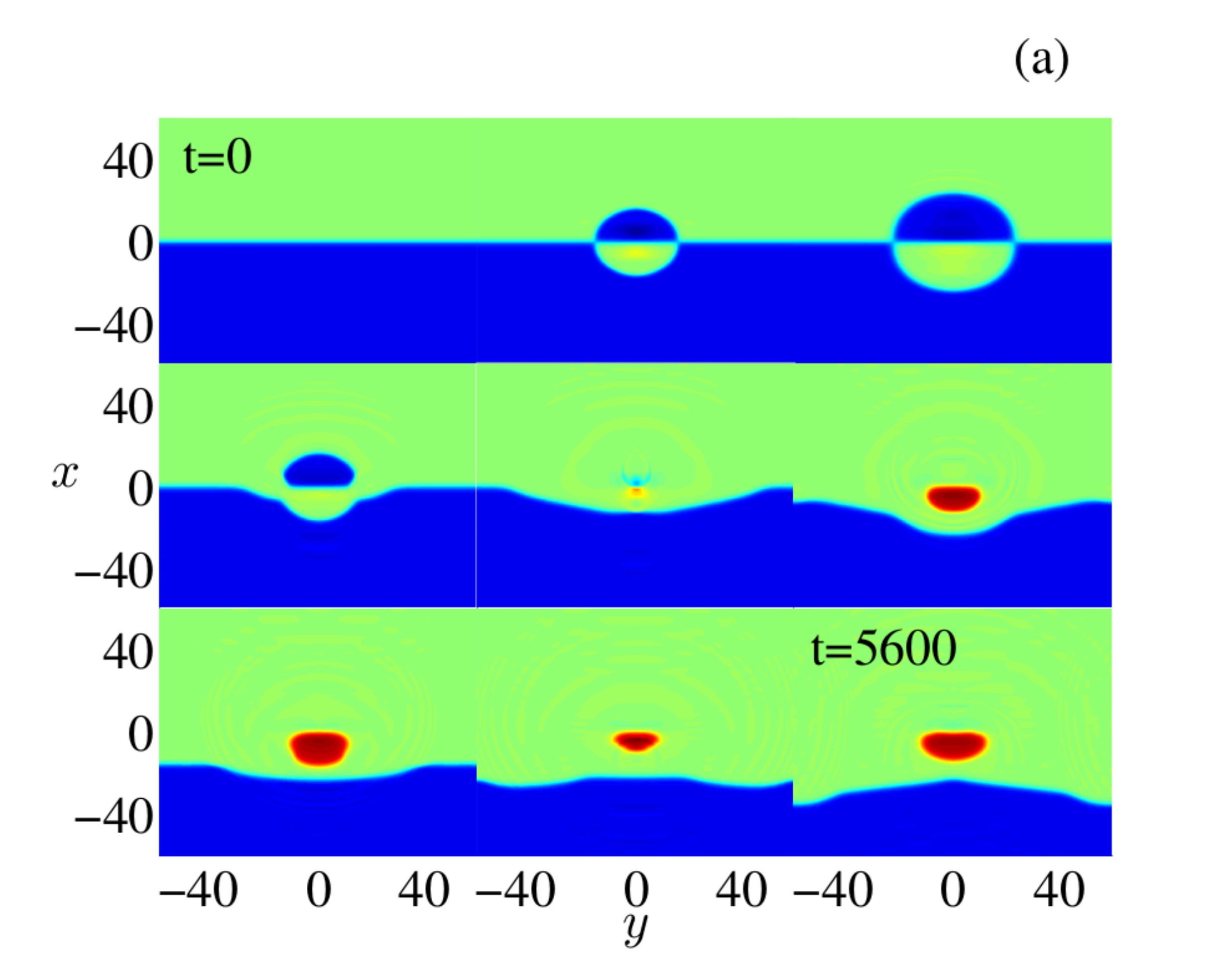}}\quad\scalebox{0.35}{\includegraphics{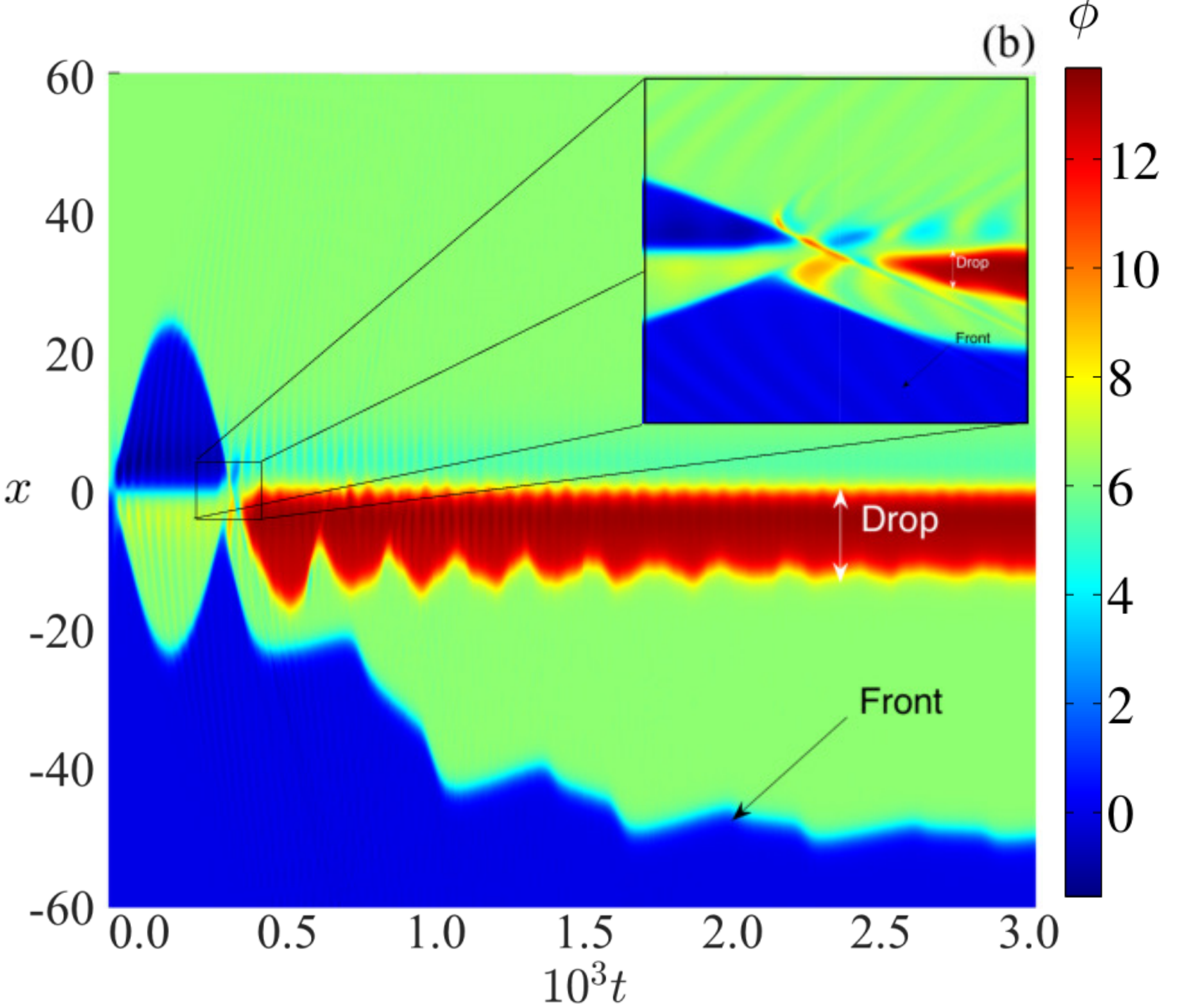}}\\
\includegraphics[width=0.95\columnwidth]{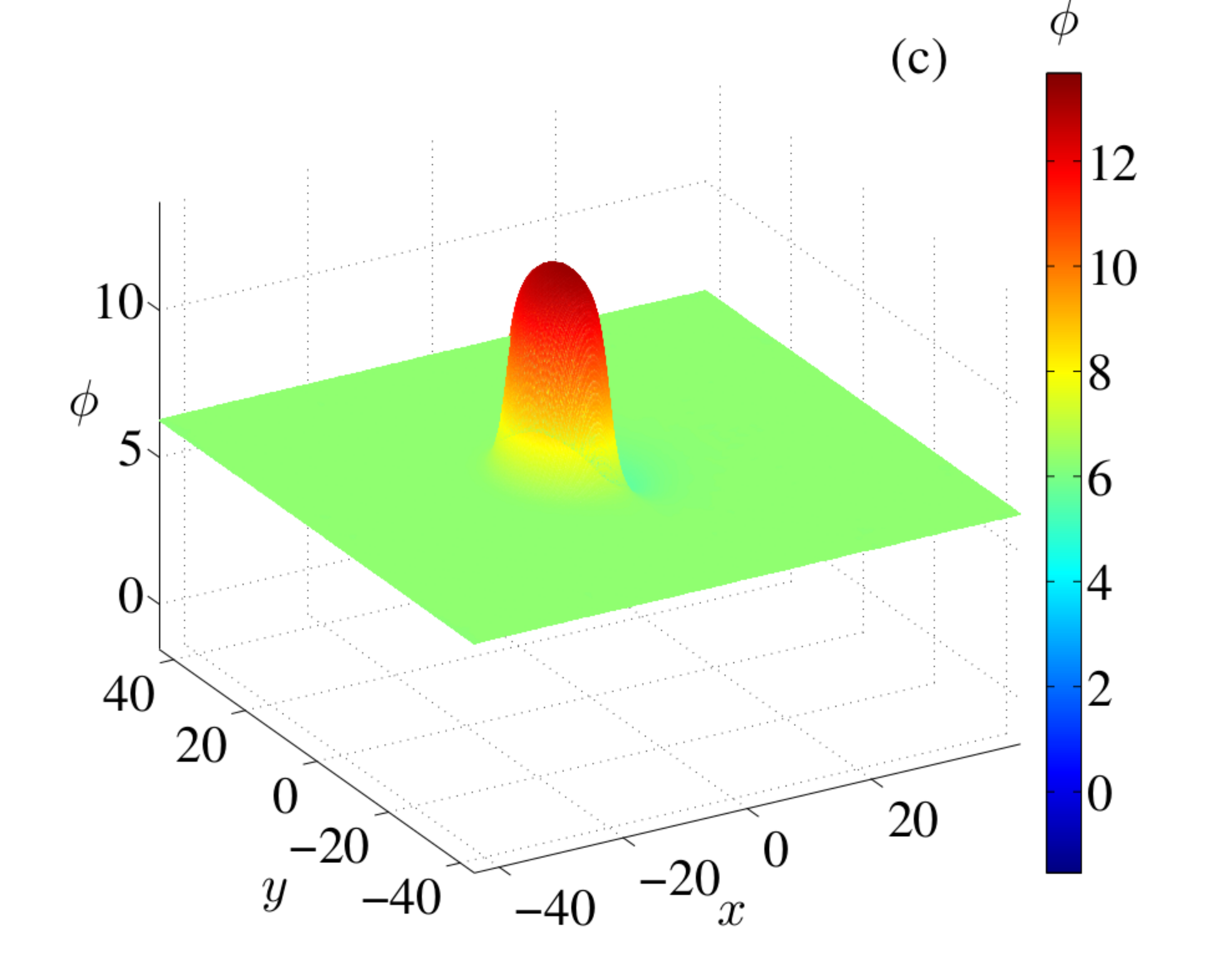}
\caption{\label{fig05} (Color online) Drop formation ($B=0.28$ and $\sigma=8.9$). (a)  Snapshots from numerical simulations, taken each 700 time
 iterations. (b)  Time evolution of the $x$-profile of the soliton for $y=0$. Inset: Zoom of the kink-antikink collision during bubble
 collapse. (c) Drop-like structure after 30000 time iterations. The value of the field $\phi$ is indicated in color scale.}
\end{figure*}

Formation and time-evolution of the bubble-like structure can also be seen in figure \ref{fig03}(b). We show the $x$-profile of the
soliton at $y=0$ as a function of time. We can see that around the first $5000$ time iterations, the elliptical-type form grows and
collapses, following by the formation of the bubble. The bubble-like structure is formed in the positive region of $x$ and its size is
initially oscillating due to reminiscent perturbations produced during its formation. These oscillations decrease in time, and the
structure remains stable with a fixed size and shape (fig. \ref{fig03}(c)). The oscillating process is depicted in Fig.\ref{fig04}. 

To evaluate the role of parameter $\sigma$ and $B$ over the formation of bubble-like structures, we proceed to vary both parameters
separately.  Figure~\ref{fig04} shows the temporal evolution of the major $R_{y}$ and minor $R_{x}$ semi-axis of the elliptical 
structure. We observe that the bubble formation process is the same in all cases but not the bubble final size. In
Fig.~\ref{fig04}(a) we have varied parameter $B$ leaving $\sigma = 15$ fixed. We can observe that both $R_{x}$ and $R_{y}$ oscillate
on time. However, after this initial transient state, the system evolves to a determined size. It is clear that as we increase $B$,
$R_{x}$ and $R_{y}$  decrease, reducing the final bubble area. On the contrary, when we vary $\sigma$, leaving $B$ fixed, the opposite
occurs. Figure~\ref{fig04}(b) shows the temporal evolution of  $R_{x}$ and $R_{y}$ of the bubble structure for $B = 0.27$ fixed. As
$\sigma$ gets smaller, we can see that the final bubble area also decreases.

However, if we continue diminishing  $\sigma$, leaving $B$ in the range where the first internal mode is unstable, the preferred structure
is one with the reverse polarity--in terms of the topological charge--a {\it drop-like structure}. Figure \ref{fig05}(a) shows the
formation of a stable drop-like structure for $B=0.28$ and $\sigma=8.9$. As we can observe, the mechanism of drop formation is similar to
that of the bubble-like structure. After the collapse of the elliptical form, a stable drop-like structure is formed in the negative
region of $x$, in contrast with the bubble formation (see Fig.~\ref{fig05}(b)). From inset in Fig.~\ref{fig05}(b) is clear that the
drop-like formation is produced after the total collapse of the initial bubble, suggesting a kink-antikink collision between the bubble
walls. Subsequently, the kink-antikink merge with the same velocity but a phase shift, enhancing a drop-like structure. In
Fig.~\ref{fig05}(c), we depicted the final stage of the drop-like structure.

From this result, we can infer that the final bubble area, and therefore $\sigma$ and $B$, are playing an important role in determining
which kind of structure will be formed, either bubble or drop-like. This will be further discussed in section \ref{Sec:Discussion}.

 \begin{figure*}[t]
\centering
\scalebox{0.32}{\includegraphics{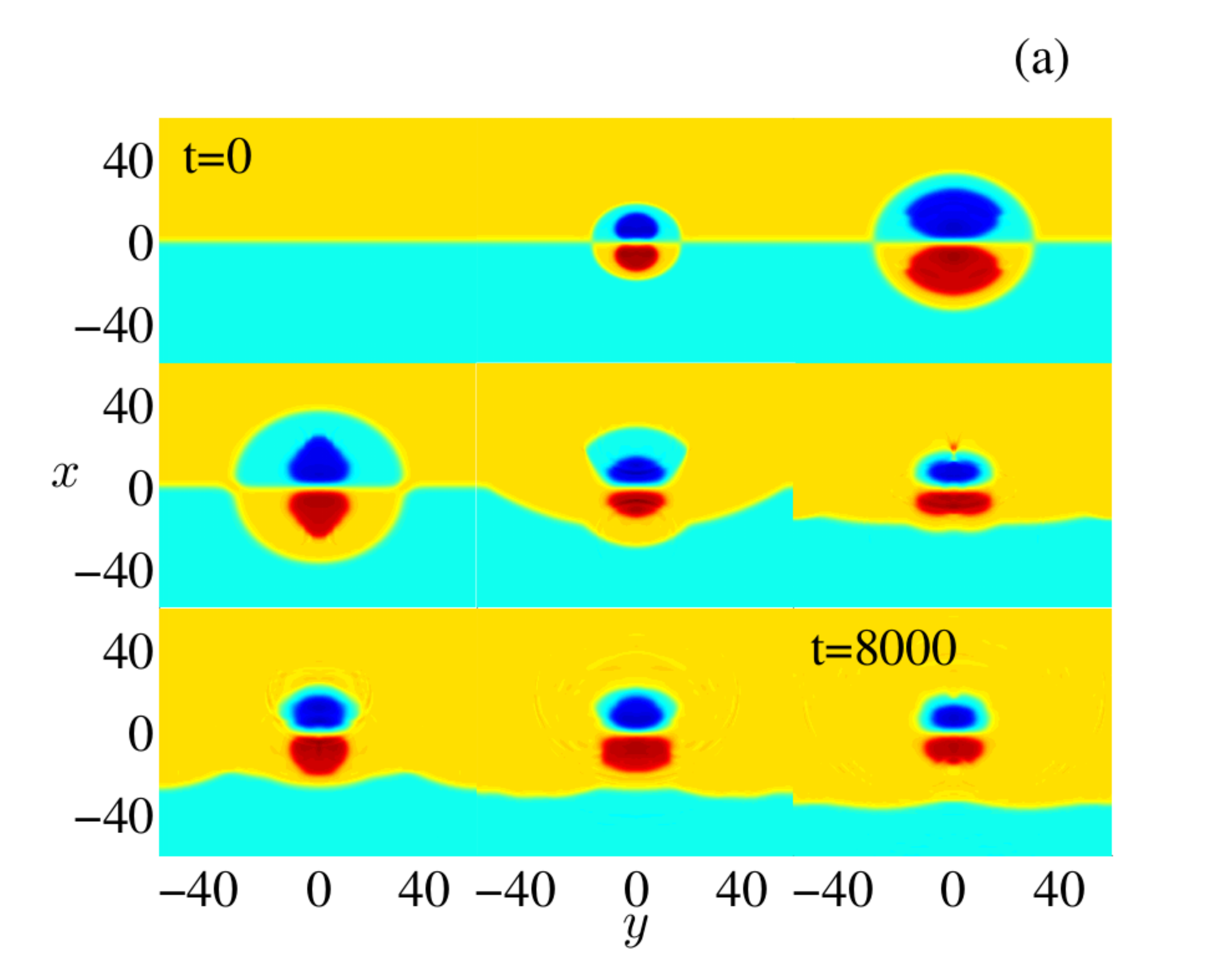}}\quad\scalebox{0.32}{\includegraphics{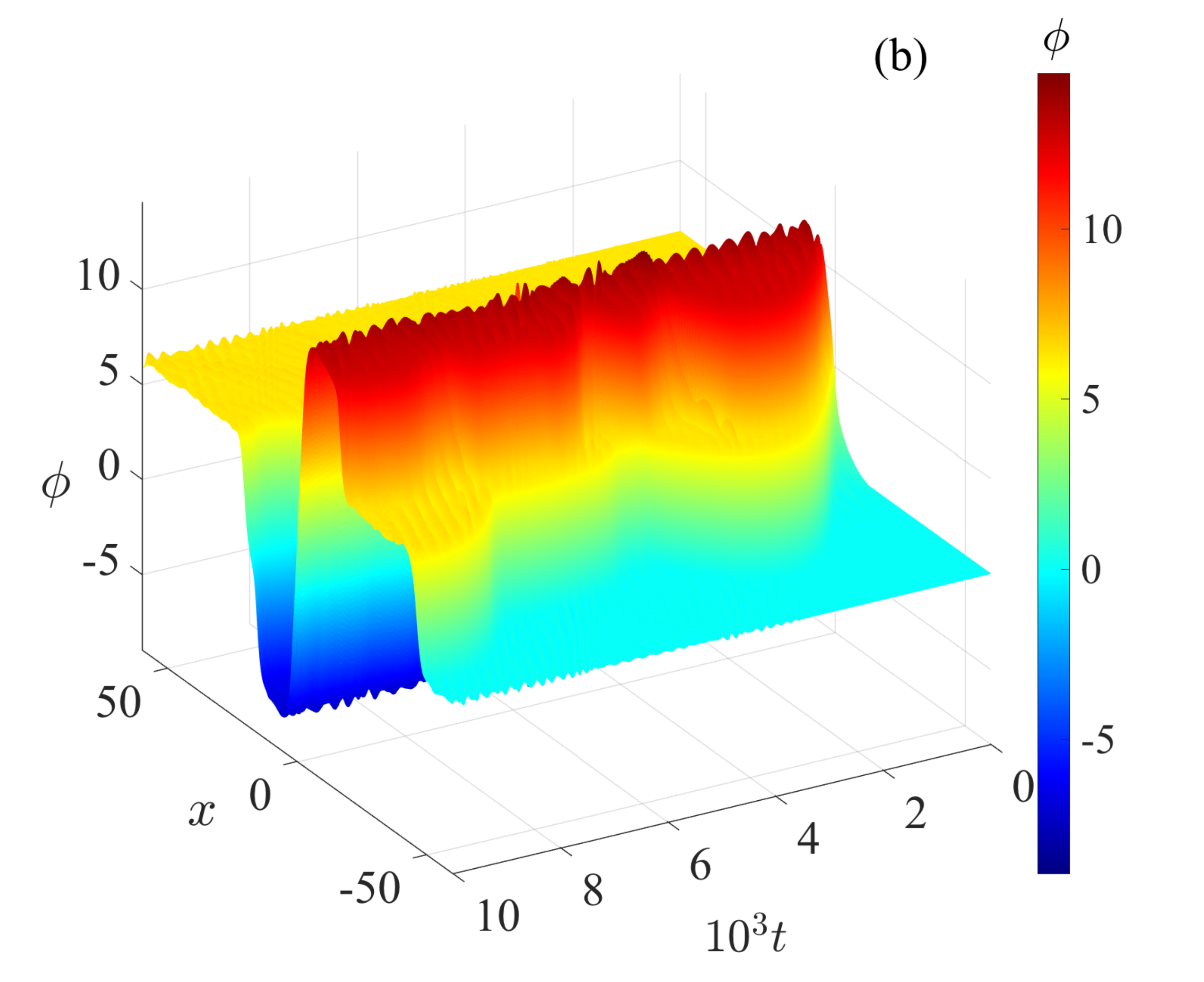}}\\
\includegraphics[width=0.95\columnwidth]{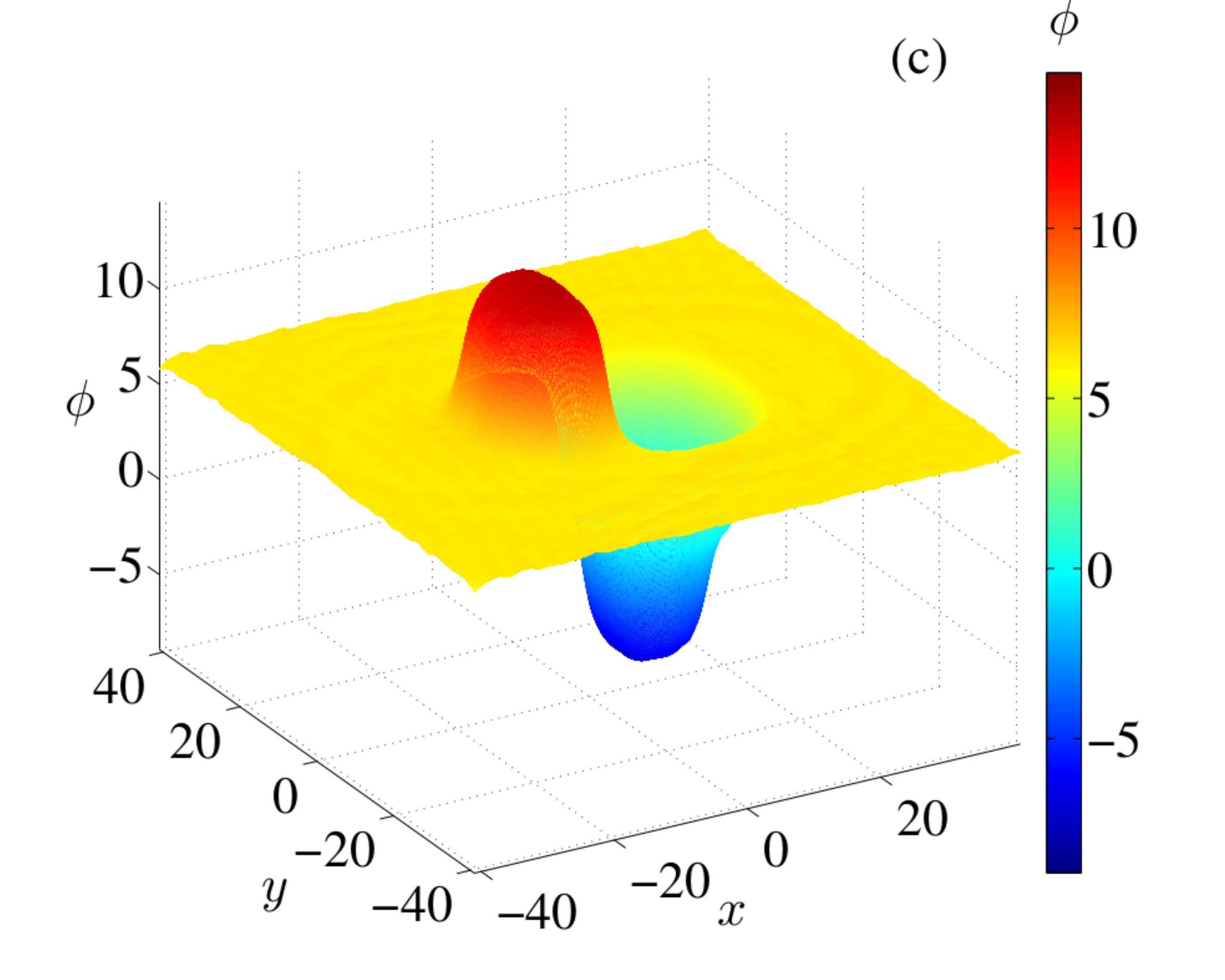}
\caption{(Color online) Pair bubble-drop formation ($B=0.2$ and $\sigma=10$). (a)  Snapshots from numerical simulations, taken each 1000
time iterations. (b) Time evolution of the $x$-profile of the soliton ($y=0$). The formation of multikinks is clearly seen. (c) Pair
bubble-drop structure after 30000 time iterations. The value of the field $\phi$ is indicated in color scale. \label{fig06}}
\end{figure*}

\subsection{Bubble-drop bound states}

Until now we have observed the formation of isolated drops or bubbles in a certain range of the parameter space $\{B, \sigma\}$. Still,
further diminishing the value of $B$, with  fixed $\sigma$, we have found a coexistence region where a bubble and a drop merge in a
stable structure--{\it a bubble-drop bound state.} In figure \ref{fig06}(a) we display the bubble-drop bound state formation for $B=0.2$
and $\sigma=10$.  From the one-dimensional theory, it is known that below $B=0.2564$, the formation of two-kink solitons occurs. As
expected, the typical elliptic-like structure grows to exhibit a two-kink profile. One two-kink connects the states $2\pi$ and $0$, and
also the states $0$ and $-2\pi$, as it can be seen in figure \ref{fig06}(b). Meanwhile, an additional two-kink connects the states $4\pi$
and $0$, passing through $2\pi$. However, after a transient, this last state breaks, the $0-2\pi$ kink moves along the negative
x-direction while the $2\pi-4\pi$ kink forms a drop. Conversely, both kinks in the positive x-direction  remain oscillating around their
positions until reach a stable state--{\it a two-kink bubble.} Finally, we get a two-kink bubble and a single drop forming a stable bound
state (see Fig.\ref{fig06}(c)). We have observed also multi-structures, like a \emph{multi-kink bubble}, for smaller values of $B$.

\begin{figure}
\scalebox{0.68}{\includegraphics{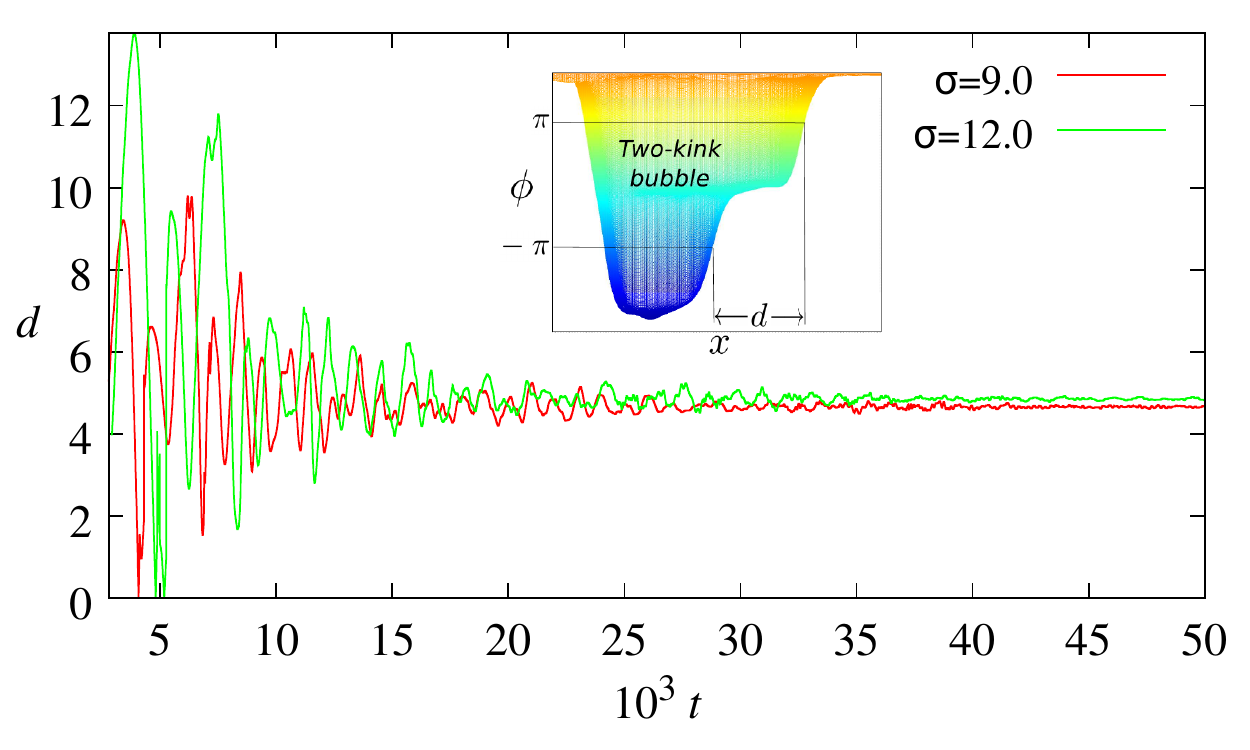}}
\caption{\label{fig07} (Color online)  Separation of kinks in a two-kink bubble as a function of time for $B=0.24$. The distance
$d$ between the cores of the kinks (inset) evolve in time before reaching a constant value. Although the final separation is a small
value compared to the total size of the system, the structure is clearly stable.}
\end{figure}

\subsection{Discussion}
\label{Sec:Discussion}

We can understand the formation of bubbles and drops by considering the following mechanism. The initial condition used in our
simulations is a $2\pi$-kink connecting two stable states in phase space, namely $\phi_0=0$ and $\phi_1=2\pi$. Since the inhomogeneous
force is localized around the origin, the internal modes of the soliton become unstable only in a neighborhood of the origin. The
instability of the internal shape mode leads to a kink-antikink pair formation. The local antikink remains steady in the origin and it will
constitute part of the bubble. This kink zone will propagate until the localized forced is weak enough to the return effect and
kink-interaction come to play. At this stage, the ellipse walls start to move backward. If $\sigma$ is large enough or $B$ small, the
walls will reach a steady position, forming a stable bubble. The final area of this bubble will depend on both parameters,
$\sigma$ and $B$.
Under a critical value of the elliptical area, the kink-antikink interaction is strong enough to produce the bubble collapse. As we have
mentioned before, this enables the formation of a drop-like structure due to a phase shift during the kink-antikink collision. Therefore,
the value of parameters $B$ and $\sigma$ will determine which kind of structure will be formed.

The external force $F(x,y)$ sustains the bubble and it acts as an opposite force of the kink-antikink interaction. The interplay
between both forces: one pulling inwards--kink-antinkink interaction--and the other pulling outwards--external force--create a stable
equilibrium-the bubble state. In fact, $F(x,y)$ is  predominant in the bubble final shape. Based on this, we can infer that $R_{x}$ is controlled
by the force \eqref{Eq02} for $y = 0$. Thus, the semi-minor axis $R_{x}$ will be proportional to the decaying coefficient of
$F_0(x)\equiv F(x,0)$, $d_{x} = 1/B$, therefore $R_{x}\sim 1/B$. Meanwhile, $R_{y}$ is controlled by the Gaussian function. The maximum of $F(x,y)$
is located at $x_{M} = 2.94$. This represents the point where the Gaussian function has its maximum extension. The decaying coefficient of
$F_{M}(y)\equiv F(2.94,y)$ is $d_{y} = 1/\sqrt{1/\sigma^2} = \sigma$. Thus,  $R_{y}\sim \sigma$. As we can see, $F_{M}(y)$ possesses a slower
decaying tail than $F_{0}(x)$, hence, the elliptical form of the bubble. Figures~\ref{fig04}(a) and~\ref{fig04}(b), respectively, corroborate these
statements.

We also notice that the shrinking force due to return effect, along with the external inhomogeneous force, produce an oscillation of the
two-kink pairs in the two-bubble structures. However, they finally stabilize forming {\it two-kink bubbles} (see figure~\ref{fig07}).

\begin{figure}
\begin{center}
 \scalebox{0.4}{\includegraphics{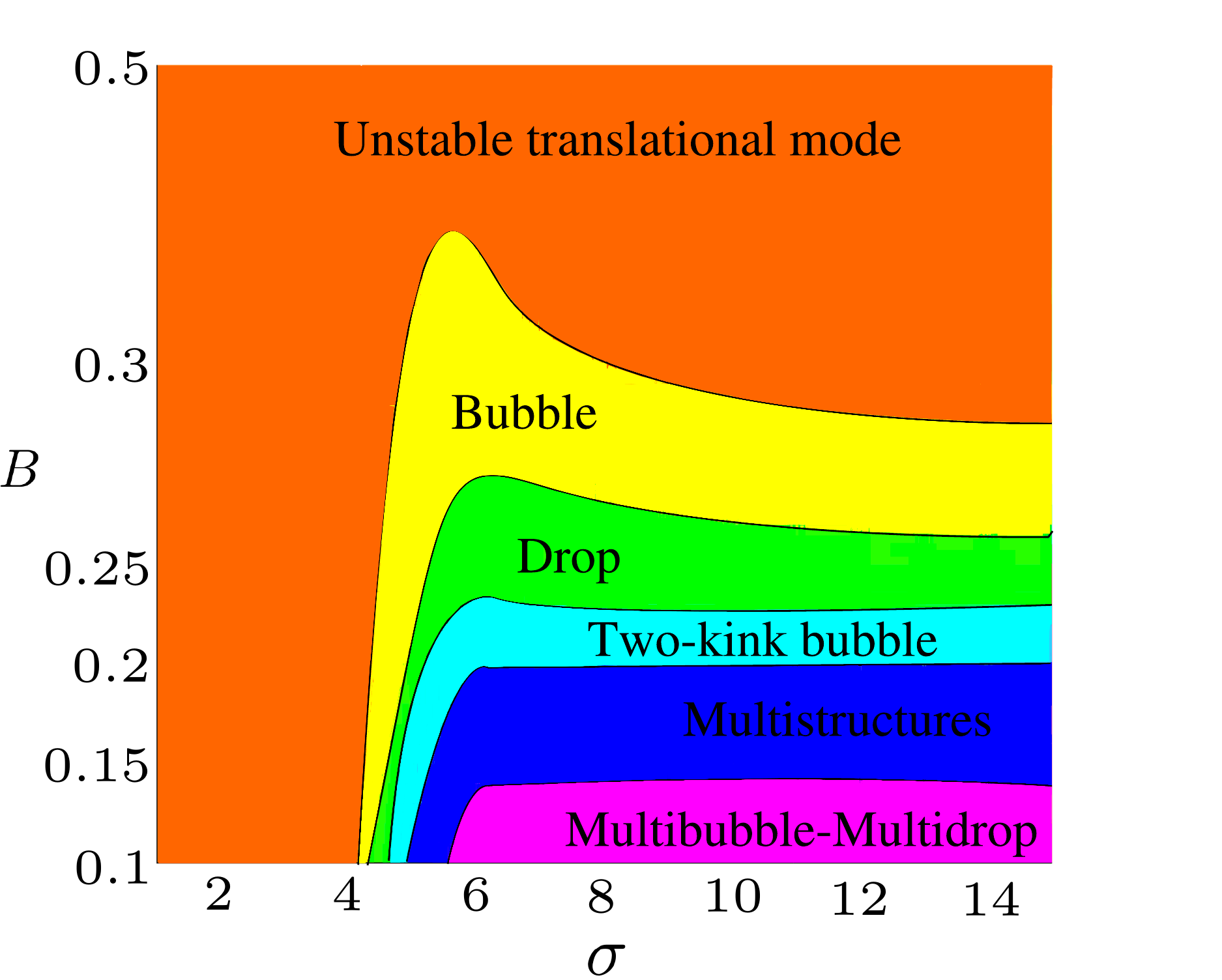}}
 \caption{(Color online) Qualitative view of the parameter space map, obtained on the basis of numerical simulations.
  Six main zones are observed, each one corresponding to different kind of structures. Bubbles and drops become more complex as increasing
  $\sigma$ and decreasing $B$.
 \label{fig08}}
 \end{center}
\end{figure}

We performed numerical simulations for several combinations of values of $B$ and $\sigma$ and obtained the qualitative parameter space map
showed in figure \ref{fig08}. All different kind of structures has been depicted. We notice that for large $\sigma$, the results are in
good agreement with the predictions of the one-dimensional theory. This is an expected result since the system behaves as a quasi-one-dimensional system in this region of parameter space.  We notice in figure \ref{fig08} that there are no stable structures below a
certain value of $\sigma$. We have calculated this minimum value $\sigma_c$ by considering the kink-antikink interaction energy,
$E_{int}(a)=-16(\sinh(a)+a)\sinh^{-1}(a)(\cosh(a)+1)^{-1}$, where $a$ is the distance between cores \cite{Bordag2012}. The kink-antikink
interaction tends to collapse the bubbles, while $F_0(x)$ tends to sustain them. Thus, we expect that stable structures can be formed when
one equilibrates the other. Indeed, the function $F_0(x)$ intersect $F_{int} = dE_{int}/da$ only in a certain interval $[\sigma_c,\infty)$.
By a careful analysis of the intersection interval, we obtained that $\sigma_c\simeq3.7$, which is in agreement with the results depicted
in figure \ref{fig08}. Under this minimum value, any structure will eventually collapse, since the inhomogeneity $F$ never equilibrates
the kink-antikink attractive force.

\section{Formation of bubbles and drops in related systems \label{Sec:4}}

It is noteworthy that the formation of bubbles and drops in our system is a robust phenomenon. Results similar to the ones
presented in this article can be obtained in other systems using other external perturbations with the same topological properties of force
\eqref{Eq02}. In the following, we briefly discuss some theoretical studies that corroborate these findings.

\subsection{Robust phenomena in dynamical systems \label{Subsec:4a}}

The sG equation \eqref{Eq01} is a particular case of the more general nonlinear KG equation
\begin{equation}
 \label{Eq03}
  \partial_{tt}\phi(\mathbf{r},t)-\nabla^2\phi(\mathbf{r},t)+\gamma\partial_t\phi(\mathbf{r},t)-G(\phi)=F(\mathbf{r}),
\end{equation}
where $G(\phi)=-\partial U(\phi)/\partial\phi$, and the nonlinear potential $U(\phi)$ possesses at least two minima separated by a barrier. These
two minima are fixed points of the associated dynamical system, and it is well known that kink solutions are heteroclinic trajectories
joining such fixed points \cite{vanSaarloos1990}. Therefore, a complete investigation of equations of type \eqref{Eq03} can be performed using the
so-called qualitative theory of dynamical systems \cite{Guckenheimer1986}. Knowing the behavior of kink solutions in the neighborhood of fixed points
and separatrices, it is possible to construct other functions with the same general properties of the exact solutions of equation \eqref{Eq03}. Thus,
it is possible to generalize the results to other equations that are topologically equivalent to those with the exact solutions \cite{Gonzalez2007}.
Therefore, we conclude that the same structure formation reported in this article can be observed qualitatively using more general forces expressed
as follows:
\begin{equation}
\label{Eq08}
F(x, y) = f(x)g(y),
\end{equation}
where $g(y)$ is an exponentially decaying function such that $\lim_{y\to\pm\infty} g(y)=0$. Thus, the general topological properties of $f(x)$ are the
following: (a) $f(x)$ should have a zero accompanied by a local minimum and a local maximum. (b) Must be localized, i.e. $\lim_{x\to\pm\infty}f(x)=0$. Note that the
function given by equation \eqref{Eq02} is a particular case of this class of functions. The relevant parameters are the slope of the function
at the equilibrium point $x = x_*$ and the values of the maximum and the minimum. 
 
To give a particular example of the above discussion, let us briefly consider in the force \eqref{Eq08} the functions
\begin{eqnarray}
\label{Eq09a}
 &f(x)=A\left[-\frac{1}{\cosh^p\left[b(x+d)\right]}+\frac{1}{\cosh^p\left[b(x-d)\right]}\right],\\
 &g(y)=\mbox{sech}\left(\frac{y}{s}\right).\label{Eq09b}
\end{eqnarray}

\begin{figure*}[t]
\centering
(a)\hspace{7.3cm}(b)
\scalebox{0.3}{\includegraphics{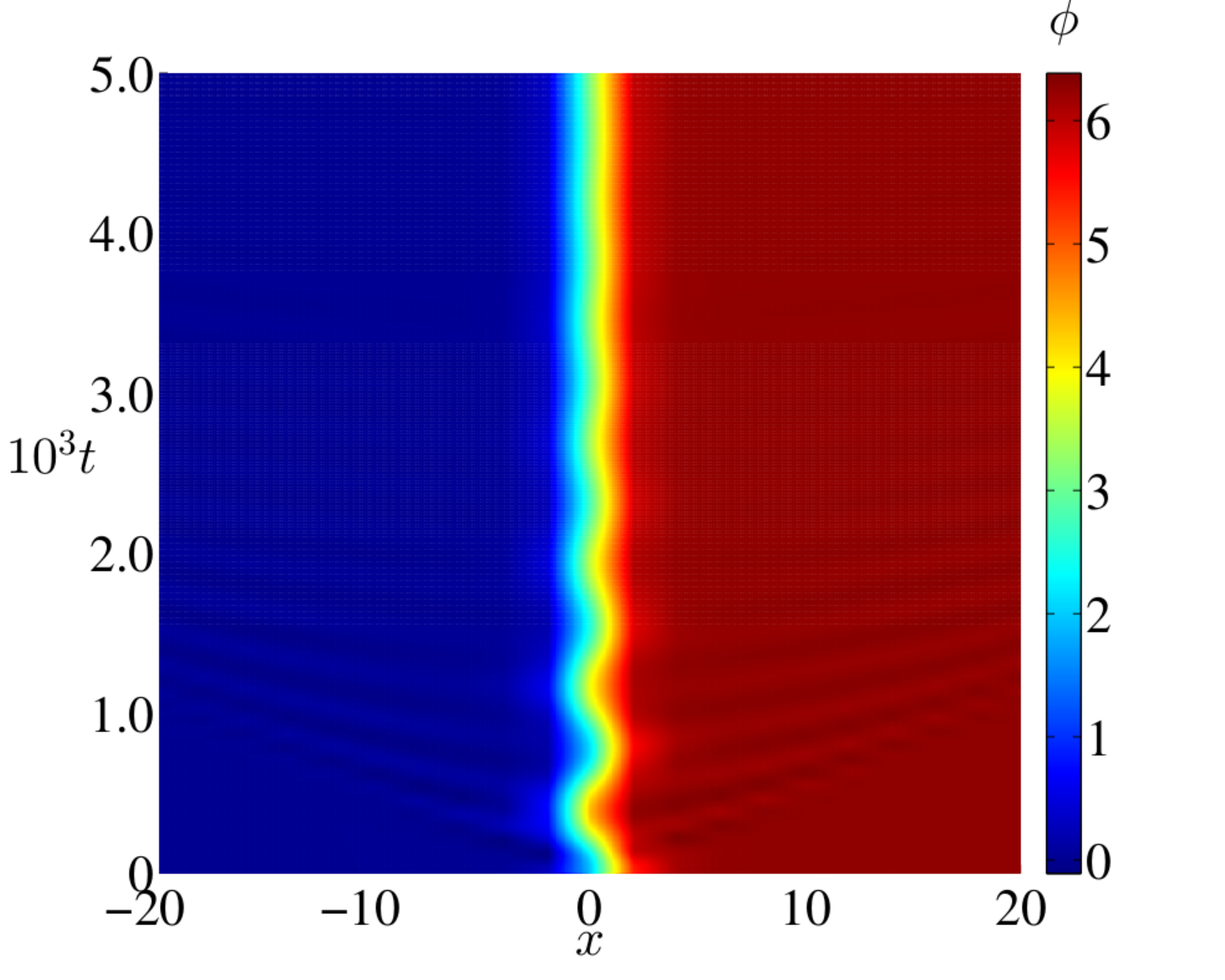}}\scalebox{0.3}{\includegraphics{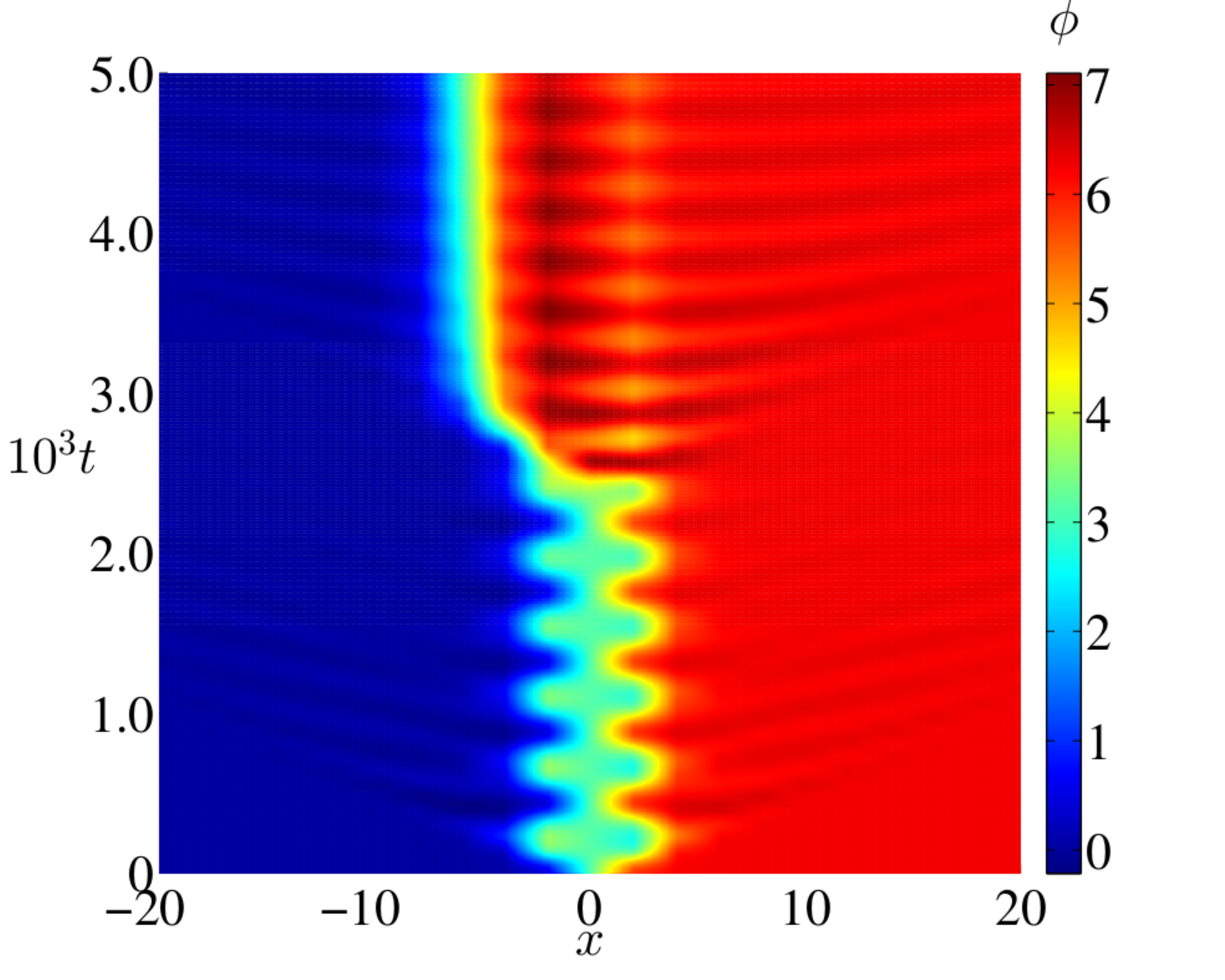}}\\
(c)\hspace{7.3cm}(d)
\scalebox{0.3}{\includegraphics{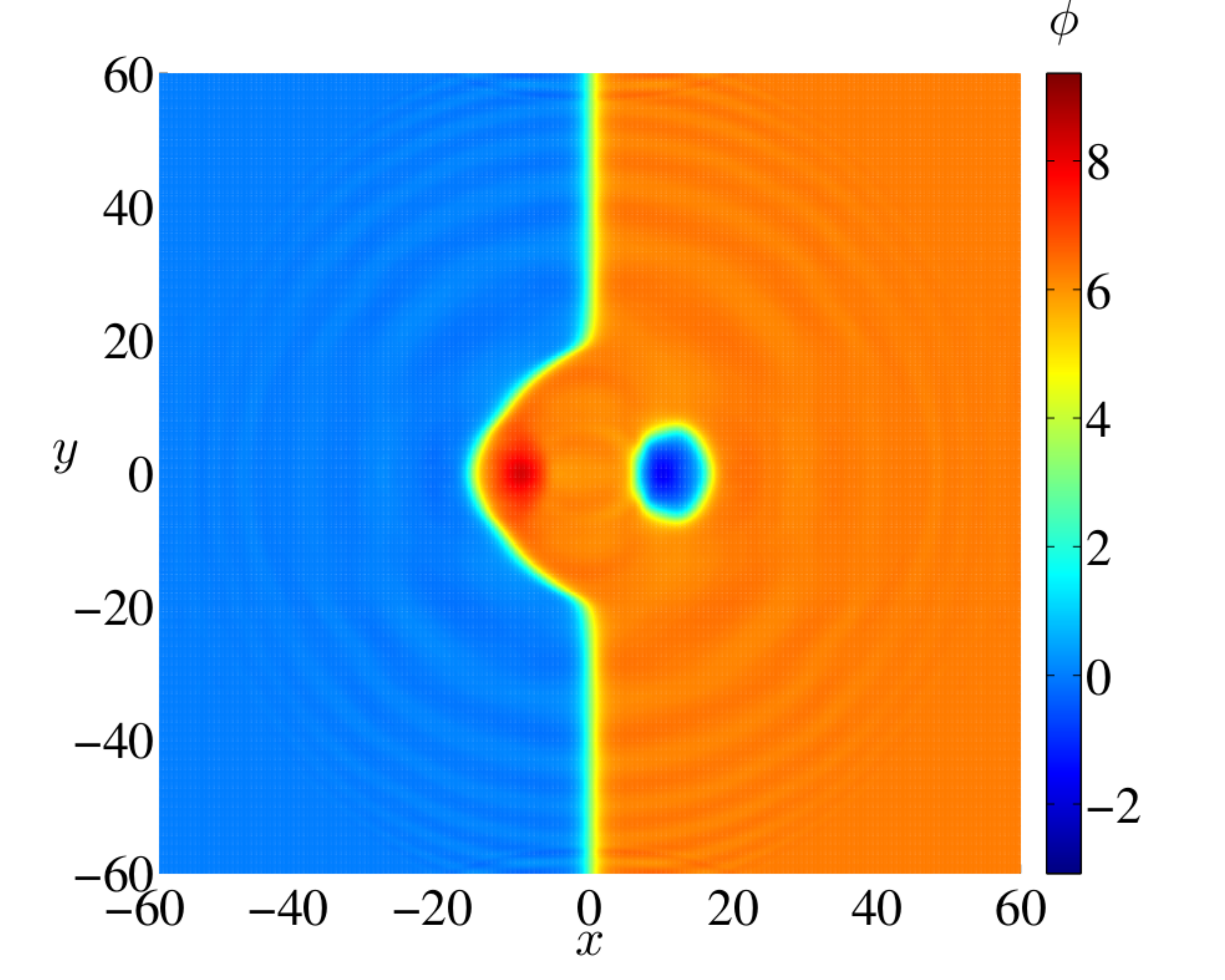}}\scalebox{0.27}{\includegraphics{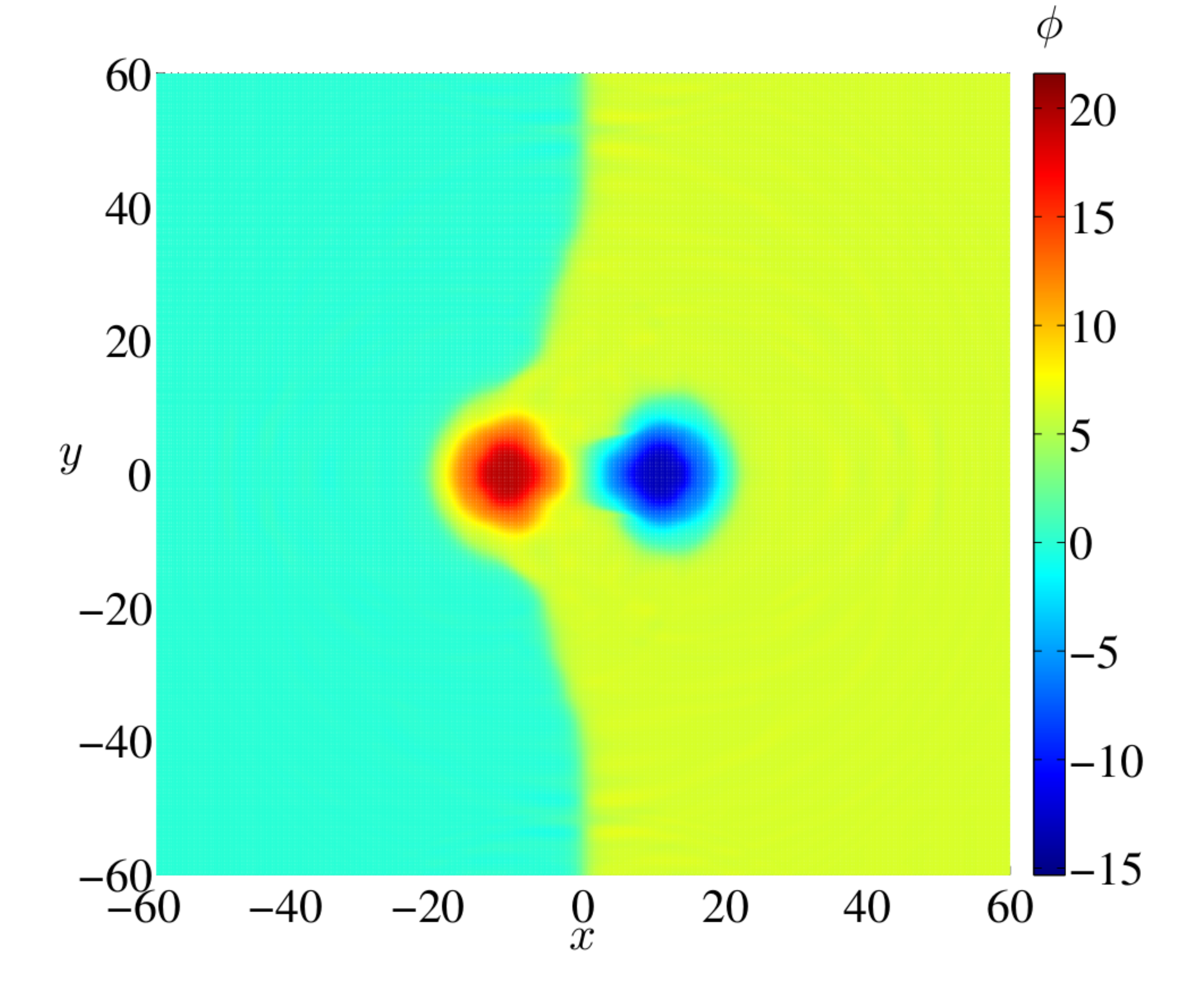}}\\
\caption{(Color online) Kink-dynamics and formation of localized structures in a topologically equivalent sG system (equations \eqref{Eq01},
\eqref{Eq08}, \eqref{Eq09a} and \eqref{Eq09b}). (a) Time evolution of the $x$-profile of an initially-at-rest kink placed at $x_o=0.6$ for
$A=1.05$, $b=1$, $d=0.5$, $p=2$ and $s=10$. The point $x_*=0$ is a stable fixed point for the kink. The soliton performs damping oscillations around $x_*$.
(b) Time evolution of the $x$-profile of a locally unstable kink for $A=-0.9$, $b=0.1$, $d=2.0$, $p=200$ and $s=0.5$. A change of stability occurs due to the change of sign in the derivative of the
force at $x_*$, and the fixed point is now unstable (see equivalent phenomenon in figure \ref{fig02}). (c) A bubble-like structure at $t=4000$ for
$A=-1.2$, $b=s=0.28$, $p=2$ and $d=10$ (see equivalent phenomenon in figure \ref{fig03}). (d) A multistructure (bound state) at $t=4000$ for $A=-1.5$,
$b=s=0.28$, $p=2$ and $d=10$ (see equivalent phenomenon in figure \ref{fig06}). \label{fig09}}
\end{figure*}

Notice that with these functional profiles, force \eqref{Eq08} fulfills all required properties.
Parameters $A$ and $d$ control the value of the local maximum of $F(x,y)$ and its $x$-coordinate position, respectively. The spread of the force in
the $x$ direction is controlled by parameters $b$ and $p$, while in the $y$-direction is controlled by $s$. In figure \ref{fig09} we reproduce qualitatively the same results
discussed in this work for the reported values of parameters.

Giving that the discussed phenomena can occur for more general forces, we conclude that $f(x)$ does not have to be exactly
$F(x) = 2(B^2 - 1) \mbox{sinh}(Bx) \mbox{sech}^2(Bx)$. For example, the $F(x,y)$ discussed in this work is a very good approximate model of the
current dipole device created by Ustinov and collaborators \cite{Ustinov2002, Malomed2004}, which can be implemented in JJ's
experiments. The technique of insertion of fluxons into annular JJ's proposed in \cite{Ustinov2002} can be theoretically understood as the creation
of a fluxon-antifluxon pair due to internal modes instabilities.
The development of methods that allow for the construction of exact solutions to the equation \eqref{Eq03} for a general
$F(\mathbf{r})$ and the investigation of the topological influence of the actual shape of the experimental forcing is an open problem that requires
more research work in the future.

\subsection{Sine-Gordon solitons in phase transitions \label{Subsec:4b}}

Equation \eqref{Eq03} also appears in the description of structural phase transitions, where a central question is whether the system can be driven
to a new phase by some instability \cite{Gonzalez2006, Gonzalez1999}.

In the vicinity of a first order phase transition, a metastable state of matter can be obtained by means of nonlinear excitations. The
main phenomenon is that of \emph{nucleation}, where bubbles and drops can grow or disappear. A \emph{critical germ} is commonly
considered for the development of a given instability that is energetically above a \emph{nucleation barrier}, in order to produce a
transition from one phase to another. If the field configuration $\phi(\mathbf{r},t)$ has a radius larger than the critical germ, the
instability will grow and a phase transition will occur \cite{Filippov1992}.

In the case of the sG system \eqref{Eq01}, the potential is given by $U(\phi)=1-\cos\phi$ \cite{Peyrard2004}, and we have several barriers, i. e., unstable
states, that separate stable states. Unstable stationary solutions that connect any two of these stable states are, indeed,
kink-antikink pairs with a drop-like or bubble-like structure. These solutions play the role of a critical germ for the transition from
one state to the other \cite{Gonzalez1989, Gonzalez1992, Gonzalez1999}. Unstable regions in parameter space can be
interpreted as intervals for parameters $\sigma$ and $B$ where the force produces drops and bubbles with a smaller size than the critical
germ, so they collapse. In the case where these structures are larger than the critical germ, the attractive force between kinks and
antikinks are balanced by the external force, so the structure is sustained by the inhomogeneity. This property has been used in real
experiments of controlled transport of bubbles using laser beams in weak absorbing liquids \cite{Gonzalez2006}, where a first-order
liquid-vapour phase transitions are induced. We expect that a time-dependent force with a spatial profile similar to equation \eqref{Eq02}
would be useful to transport the bubble-like and drop-like structures reported in this work.

\section{Conclusions and final remarks \label{Sec:6}}

By means of numerical simulations, we have shown that the internal structure of two-dimensional sine-Gordon solitons affects the dynamics
of the wave when interacting with localized inhomogeneities. We have observed the formation of bubble and drop-like structures that are
sustained by the force, being unstable otherwise. The formation of these structures are entirely due to internal mode instabilities of the
solitary wave, and it has no relation with any scattering effect. We have given a qualitative explanation of the formation of such structures
using an analytical model of the breaking of solitons by topologically equivalent inhomogeneities. From our simulations, we have
demonstrated that the two-kink solutions can be also stable in two-dimensional systems. For the observed phenomena, we have given an
interpretation in the context of the phase transitions theory. We expect to observe the formation of bubbles and drops
in related solitonic systems, like the $\phi^4$ model. Of course, the existence of multi-structures will depend on the existence of many
stable states in phase space. Thus, we do not expect to find a two-kink bubble structure in a two-dimensional  $\phi^4$ model, as well as
it is not expected to find multikinks in the one-dimensional model \cite{GarciaNustes2012}.

As a final remark, the phenomena observed in this work may have important implications in two-dimensional JJ's devices. The
formation of bubbles in the junction due to a dipole current and the trapping and breaking of such structures by such localized inhomogeneities
may produce important effects on the functioning of the device.  Moreover, the study of the stability and transport of such bubbles can shed light
on the design of built-in heterogeneities and control of perturbations in the junction. Fluxon transmission in JJ's devices can also be
regulated by the application of this kind of external perturbations.

\begin{acknowledgments}
 The authors thank Irving Rond\'on for fruitful discussions. M.A.G-N. thanks for the financial support of FONDECYT project 11130450. J.F.M.
 acknowledge financial support of CONICYT doctorado nacional 21150292.
\end{acknowledgments}

\providecommand{\noopsort}[1]{}\providecommand{\singleletter}[1]{#1}%

\end{document}